\documentclass[12pt,twoside,preprint,prd]{revtex4}
\usepackage{amsmath,amsfonts}

\newcommand{\bea}{\begin{eqnarray}}
\newcommand{\eea}{\end{eqnarray}}

\newcommand{\rd}[1]{\mathop{\mathrm{d}#1}}
\newcommand{\rD}[1]{\mathop{\mathrm{D}#1}}

\newcommand{\fract}[2]{{\textstyle\frac{#1}{#2}}}
\newcommand{\grad}{\vec\nabla}

\newcommand{\vj}{{\vec j}}
\newcommand{\vk}{{\vec k}}
\newcommand{\vcr}{{\vec r}}
\newcommand{\vv}{{\vec v}}

\newcommand{\vrr}{{\vec r}}
\newcommand{\vf}{{\vec f}}
\newcommand{\vx}{{\vec x}}
\newcommand{\vchi}{{\vec \chi}}

\newcommand{\ha}{\hat A}

\newcommand{\han}[1]{\hat A_{#1}}
\newcommand{\vha}{\skew3\hat{\vec A}}

\newcommand{\tenshf}{\skew2\hat{\mathsf F}}

\newcommand{\vD}{{\vec D}}

\newcommand{\vX}{{\vec X}}

\newcommand{\pr}{\partial}

\newcommand{\paj}{\partial_j}

\newcommand{\mn}{{\mu\nu}}
\newcommand{\tvr}{(t,\vrr)}
\newcommand{\tvx}{(t,\vx)}
\newcommand{\tvX}{(t,\vX)}
\newcommand{\tvk}{(t,\vk)}

\newcommand{\fvx}[1]{f^{#1}(\vx)}
\newcommand{\tn}[2]{\theta^{#1#2}}
\newcommand{\en}[2]{\eps^{#1#2}}
\newcommand{\on}[2]{\omega_{#1#2}}
\newcommand{\vom}{\vec\omega}

\newcommand{\vth}{\vec\theta}
\newcommand{\tensth}{\vec\theta}
\newcommand{\pa}[1]{\frac\partial{\partial #1}}
\newcommand{\pas}[1]{\partial_{#1}}

\newcommand{\co}{coordinate}

\newcommand{\rk}{{\rm k}}
\newcommand{\rX}{{\rm X}}
\newcommand{\Tr}{{\rm Tr}}
\newcommand{\rJ}{{\rm J}}

\newcommand{\fk}{{\mathfrak f}}
\newcommand{\fkf}{{\vec{\mathfrak f}}}

\renewcommand{\Re}{\operatorname{Re}}
\renewcommand{\Im}{\operatorname{Im}}

\newcommand{\numeq}[2]{\begin{equation}
#2
\label{#1}
\end{equation}}
\newcommand{\rfq}[1]{~(\ref{#1})}
\newcommand{\citer}[1]{~\cite{#1}}

\let\vec\boldsymbol
\let\eps\varepsilon
\let\epsilon\varepsilon
\let\hat\widehat

\newcommand{\ct}{coordinate transformation}
\newcommand{\df}{diffeomorphism}

\newcommand{\nc}{noncommutative}
\newcommand{\ncg}{noncommuting}
\newcommand{\sw}{Seiberg-Witten}

\numberwithin{equation}{section}
\begin{document} 

\title{Noncommuting Gauge Fields as a Lagrange Fluid}

\author{R. Jackiw}
\affiliation{Center for Theoretical Physics\\
Massachusetts Institute of Technology\\
Cambridge, MA 02139-4307, USA}
\author{S.-Y. Pi}
\affiliation{Physics Department, Boston University \\
Boston, MA 02215, USA}
\author{A.P. Polychronakos}
\affiliation{Physics Dept., Rockefeller University\\
New York, NY 10021-6399, USA\\{\rm and}\\
Physics Dept.,  University of Ioannina\\ 45110
Ioannina, Greece}

\vskip 0.2in
 
\date{1 June  2002\\
\null}

\begin{abstract}\noindent
The Lagrange description of an ideal fluid gives rise in a natural way 
to a gauge potential and a Poisson structure that are classical precursors 
of analogous  noncommuting entities.  With this observation we are led to 
construct gauge-covariant coordinate transformations on a noncommuting space. 
Also we recognize the Seiberg-Witten map from noncommuting to commuting 
variables as the quantum correspondent of the Lagrange to Euler map 
in fluid mechanics.
\end{abstract}

\preprint{MIT-CTP-3263}
\preprint{BUHEP-02-18}
\preprint{RU-02-3-B}

\maketitle

\markboth{\small {\it R. Jackiw, S.-Y. Pi, and A.P. Polychronakos}}{\small  Noncommuting
Gauge Fields}

\section{Introduction}\label{s1}

Noncommuting coordinates are characterized by a constant, antisymmetric 
tensor $\tn ij$:
\numeq{1.1}{
[x^i, x^j] = i\tn ij\ .
}
Subjecting the \co s to an infinitesimal \co\ transformation
\numeq{1.2}{
\delta \vec x = - \vec f(\vx)
}
and requiring that\rfq{1.1} remain unchanged results in the condition
\numeq{1.3}{
-[\fvx i, x^j] -[x^i, \fvx j] = 0
}
which in turn implies  by\rfq{1.1} that 
\numeq{1.4}{
-\pas k \fvx i \tn kj - \pas k  \fvx j \tn ik  = 0\ .
}
The left side is recognized as the Lie derivative of a contravariant tensor
\numeq{1.5}{
L_f \tn ij = f^k \pas k \tn ij - \pas k  f^i \tn kj - \pas k f^j \tn ik
}
with the first term on the right vanishing since $\vth$ is constant. So the
\nc\ algebra\rfq{1.1} is preserved by those coordinate transformations that 
leave $\vth$ invariant: $L_f \vth = 0$.

To unravel the  condition\rfq{1.4}, we must specify whether $\vth$ possesses an
inverse~$\vom$:
\numeq{1.6n}{
\tn ij \on jk = \delta^i_k\ . 
}
An inverse can exist in even dimensions, provided $\vth$ is nonsingular, but $\vom$ will
not exist in odd dimensions, where the antisymmetric $\vth$ always possesses a zero
mode. We shall assume the generic situation: nondegenerate $\vth$ with no zero modes
in even dimensions, one zero mode in odd dimensions.

To solve for $\vf$ in even dimensions, we define
\numeq{1.7n}{
f^i = \tn ij g_j\quad (i,j=1,\ldots,2n)\ .
}
This entails no loss of generality, because $\vth$ is nonsingular (by hypothesis).
Then\rfq{1.4} becomes
\begin{subequations}\label{3.1}
\numeq{3.1a}{
\tn i\ell \pas k g_\ell  \tn kj + \tn j\ell \pas k  g_\ell \tn ik  = 0\ .
}
Because $\vth$ is nonsingular and antisymmetric, this implies
\numeq{3.1b}{
 \pas k g_\ell  - \pas \ell  g_k  = 0 
}
\end{subequations}
or
\numeq{3.2}{
  g_\ell  = \pas\ell\phi\ .
}
Thus we have 
\numeq{3.3}{
f^i = \tn ij \pas j  \phi 
}
for the \ct s (in even dimensions) that leave $\vth$ invariant. Since
\numeq{1.11}{
\grad\cdot \vf = 0\ .
}
the transformations are volume preserving; the  Jacobian of the finite 
\df\ is unity.  However, except in two dimensions, these are not the most 
general volume-preserving transformations.  Nevertheless, they form a group: 
the Lie bracket of two transformations like\rfq{3.3}, 
$f_1^i = \tn ij \paj\phi_1$ and $f_2^i = \tn ij \paj\phi_2$,
takes the same form, $\tn ij\paj(\tn k\ell \pr_k\phi_1 \pr_\ell \phi_2)$.  
The group is the symplectic subgroup of volume-preserving \df s that also 
preserve~$\tn ij$. 

In two dimensions, where we can set $\tn ij=\theta \en ij$, the above 
transformations exhaust all the area-preserving transformations. 

In odd dimensions, where (by assumption) $\vth$ possesses 
a single zero mode, for definiteness we orient the
coordinates so that the zero mode lies in the first direction (labeled 
$0\to$~time) and
$\vth$, confined to the remaining (spatial) dimensions, is nonsingular:
\numeq{1.12}{\begin{aligned}
\theta^\mn &= \biggl(\begin{matrix} 
0 & 0\\
0 &\tn ij
\end{matrix} \biggr) \quad  (i,j = 1,\ldots,2n) \\
\tn ij \on  jk &= \delta_k^i\ .
\end{aligned}
}
The \df s that preserve $\vth$ now take the infinitesimal form
\numeq{1.13}{
f^\mu = \Biggl\{ \begin{array}{l}
f(t) \\
\displaystyle 
\tn ij \pa{x^j} \phi\tvx
\end{array} 
}
These still form a group. Two transformations, 
$(f_1,\phi_1)$ and $(f_2,\phi_2)$,
possess a  Lie bracket of the same form\rfq{1.13}, with 
$(f_2\pr_t f_1 - f_1\pr_t f_2, f_2\pr_t\phi_1 - f_1\pr_t\phi_2 
+ \tn k\ell \pr_k \phi_1\pr_\ell\phi_2)$. 
But the space-time volume is not preserved: $\pr_\mu f^\mu \neq
0$. (Of course, at fixed time, the spatial volume is preserved.)

Unit-Jacobian diffeomorphisms also leave invariant the equations 
for an ideal fluid, in the Lagrange formulation of fluid mechanics, 
and in particular a planar (two dimensional) fluid
supports area-preserving diffeomorphisms. This coincidence of invariance 
suggests that other aspects of noncommutativity possess analogs in the 
theory of fluids,
whose familiar features can therefore clarify some obscurities of
noncommutativity. (A similar point of view was taken by Susskind\citer1 in the
description of the quantum Hall effect.)

In this paper we explore connections between fluid mechanics and \ncg\ field
theory, first in Section~\ref{s2} for low-dimensional systems, and then in
Section~\ref{s3} for higher-dimensional theories. The natural Poisson 
(commutator) structure, present in the Lagrange description of a
fluid, and the possibility of introducing a vector potential to describe 
the evolution of comoving coordinates, will be recognized as classical 
precursors of analogous
noncommuting entities. Within this framework, we shall show how noncommuting 
gauge fields respond to coordinate transformations, generalizing previously
established results\citer2. Also we shall demonstrate that the \sw\ map between
\ncg\ and commuting gauge fields\citer3 corresponds to the mapping between 
the Lagrange and Euler formulations of fluid mechanics. 
In this context it is possible to rederive simply the explicit ``solution'' 
to the \sw\ map in even dimensions\citer4 and to
extend it to odd dimensions.

We conclude this Introduction by recalling the two formulations of fluid dynamics\citer5.
The Lagrange description uses the coordinates of the particles comprising the fluid:
 $\vX \tvx$.  These are labeled by a set of parameters~$\vx$, which are the
coordinates of some initial reference configuration, e.g., $\vX(0,\vx) = \vx$, and are called
comoving coordinates. 
We may parameterize the evolution of $\vX$ by defining
\numeq{1.14}{
X^i \tvx = x^i + \tn ij \han j \tvx
}
which loses no generality provided $\vth$ is nonsingular. As will be seen below, $\vha$
behaves as a noncommuting, Abelian vector potential. 

In the  Euler description $\vX$ is
promoted to an independent variable and renamed~$\vcr$.  Dynamics is described by 
the space-time--dependent density $\rho\tvr$ and
velocity $\vv\tvr$. 
The two formulations are related by postulating sufficient regularity so that 
(single-valued) inverse
functions exist:
\numeq{1.15}{
\vX \tvx \Bigr|_{\vx = \vchi\tvr}\!\! =  \vcr
}
$\vX \tvx$ provides a mapping of the original position $\vx$  to position at time~$t$:
$\vX=\vcr$, while $\vchi\tvr$ is the inverse mapping.
 The Eulerian density then is defined
by
\begin{subequations}\label{1.16}
\numeq{1.16a}{
\rho\tvr =  \int \rd x \rho_0(\vx) \delta \bigl(\vX \tvx - \vcr\bigr)\ .
}
where $\rho_0(\vx)$ is a reference density, usually taken to be homogeneous:
\numeq{1.16b}{
\rho\tvr = \rho_0 \int \rd x \delta\bigl(\vX \tvx - \vcr\bigr)\ .
}
\end{subequations}
(The integral and the $\delta$-function carry the dimensionality of the relevant
space.) Evidently this evaluates as 
\numeq{1.17}{
\frac1{\rho\tvr} = \frac1{\rho_0} \det \frac{\pr X^i \tvx}{\pr x^j}\Bigr|_{\vx =
\vchi\tvr} \ . 
}
The Eulerian velocity is 
\numeq{1.18}{
\vv\tvr =  \dot{\vX} \tvx \Bigr|_{\vx =\vchi\tvr}  
}
where the overdot denotes differentiation with respect to the explicit time dependence.
[Evaluating an expression  at $\vx=\vchi\tvr$ is equivalent to eliminating $\vx$ in favor of
$\vX$, which is then renamed~$\vcr$.]
It is also true that the current $\vj=\rho\vv$, given in terms of Lagrange variables
by 
\numeq{1.19}{
\vj\tvr = \rho_0 \int \rd x \dot{\vX} \tvx\, \delta\bigl(\vX\tvx - \vcr\bigr) 
}
obeys a continuity equation as a consequence of the above definitions:
\numeq{1.20}{
\dot\rho + \grad \cdot \vj = 0\ .
}

The kinetic part of the Lagrangian for the Lagrange variables is simply
\numeq{1.21}{
L_0 =  \int \rd x \fract12 \dot{\vX} \tvx \cdot  \dot{\vX} \tvx\ .
}
This is invariant against the infinitesimal diffeomorphism\rfq{1.2}, provided $\vX$
transforms as a scalar
\numeq{1.22}{
\delta\vX = \vf \cdot \grad\vX
}
and $\vf$ is transverse,\rfq{1.11}. When the interaction Lagrangian is taken as
\numeq{1.23}{
L_I = - \int \rd x  V\Bigl(\det \frac{\pr X^i \tvx}{\pr x^j}\Bigr) 
}
its variation under\rfq{1.22}, with transverse $\vf$, also vanishes so that
volume-preserving diffeomorphisms remain symmetries of the interacting theory. These
are not symmetries of dynamics; rather they describe redundancy in the description: the
transformations\rfq{1.2},\rfq{1.22} relabel the parameters~$\vx$; in a sense, which is
made precise below, they are gauge transformations. 

Although we shall not need this, we note for completeness that the equation of motion for
the Lagrange variables 
\numeq{1.24}{
\ddot{X}^i \tvx = \pa{x^j}  \Bigl[  \Bigl(\frac{\pr X^i \tvx}{\pr x^j}\Bigr)^{-1}
  V'\Bigl(\det \frac{\pr X^k \tvx}{\pr x^\ell}\Bigr)
 \det \frac{\pr X^m \tvx}{\pr x^n} \Bigr]
}
implies that the Euler velocity $\vv$ satisfies an evolution equation, which follows by
differentiating\rfq{1.19} with respect to time, and using\rfq{1.18},
\rfq{1.20} and \rfq{1.24}, 
\numeq{1.25}{
\dot\vv + \vv\cdot\grad\vv = - \frac1\rho \grad P(\rho)\ .
}
Here the pressure $P$ is given by
\begin{align}
P(\rho) &= -\int \rd x V' \Bigl(\det \frac{\pr X^k \tvx}{\pr x^\ell}\Bigr)
\det \frac{\pr X^i \tvx}{\pr x^j} 
\delta\bigl( \vX \tvx  - \vcr\bigr)
\notag \\
&= -V'\Bigl(\frac1\rho\Bigr)\ .\label{1.26}
\end{align}
(The prime denotes derivation with respect to argument.)

By multiplying $L_0 + L_I$ by unity in the form $\int\rd r \delta\bigl(
\vX\tvx - \vcr\bigr)$ and performing the  $\vx$~integral with the help of\rfq{1.16}
and\rfq{1.19} we obtain a Lagrangian in terms of Euler variables:
\numeq{1.27}{
L = \frac1{\rho_0} \int\rd r \biggl(\fract12 \rho\vv^2 - \rho
V\Bigl(\frac1\rho\Bigr)\biggr) 
\ . }
The  Euler variables $\rho, \vv$ do not change under the relabeling symmetry\rfq{1.22}
of the Lagrange parameters.  Since these parameters are absent in the Euler formulation,
these diffeomorphisms are invisible. 

\section{\!Noncommuting Gauge Theory (Primarily in Low Dimensions)}\label{s2}

\subsection{Commuting theory with Poisson structure}\label{s2ssA}
We introduce into the Lagrange fluid description the (nonsingular) antisymmetric tensor
$\vth$. This allows for a natural definition of a  Poisson bracket, which may be viewed as a
classical precursor of the noncommutativity of coordinates. 
We define the bracket by 
\numeq{2.2a}{
\{\mathcal O_1,\mathcal O_2\} = \tn ij \frac{\pr \mathcal O_1}{\pr x^i} \frac{\pr \mathcal
O_2}{\pr x^j} }
so that
\numeq{2.2b}{
 \{ x^i, x^j\} = \tn ij\ .
}
It follows from the definition\rfq{1.14} that
\numeq{2.4a}{ 
\{X^i,X^j\}  = \tn ij + \tn ik \tn j\ell \hat F_{k\ell} 
}
with
\numeq{2.5}{
\hat F_{ij} = \pa{x^i} \ha_j - \pa{x^j} \ha_i + \{\ha_i, \ha_j\}\ .
} 

It is seen that the structure of the gauge field $\tenshf$ is as in a noncommuting theory,
with the Poisson bracket replacing the commutator of two potentials~$\vha$.
Also, in the limit that the deviation of $\vX$  from the reference configuration~$\vx$ is
small, that is, for small~$\vha$, we recover a conventional Abelian gauge field. 

The above formulas are understood to hold either in even dimensions for a purely spatial
Euclidean formulation (there is no time variable) or in odd-dimensional space-time for
 spatial components ($\vX$ and $\vx$ are spatial vectors, without time components). 

\subsection{Commuting transformations (even dimensions)}\label{s2ssB}
In even dimensions, the $\vth$-preserving transverse \df, which also implements the
reparameterization symmetry of the Lagrange fluid, acts on $\vX$  through the bracket
[see\rfq{3.3},\rfq{1.22}, and\rfq{2.2a}]:
\numeq{2.7}{
\delta_\phi \vX = \tn ij \frac{\pr \vX}{\pr x^i}  \frac{\pr\phi(\vx)}{\pr x^j} 
= \{ \vX, \phi(\vx)\} \ .
}
Because $\delta \vX$ compares the transformed and untransformed $\vX$
at the same argument,
$\delta \han i = \on ij \delta X^j$ and the volume-preserving
diffeomorphism\rfq{2.7} induces a gauge transformation on $\vha$: 
\begin{subequations}\label{2.8}
\begin{align}
\delta_\phi \vha(\vx)  &= \grad\phi(\vx) + \{\vha(\vx), \phi(\vx)\} 
       \equiv \vD \phi \label{2.8a}\\
\delta_\phi \hat F_{ij}(\vx) &= \{\hat F_{ij} (\vx), \phi(\vx)\}\ .\label{2.8b}
\end{align}
\end{subequations} 
We see that the dynamically sterile relabeling diffeomorphism of the parameters in the
Lagrange fluid leads to an equally sterile gauge transformation, under which $\vX$ and
$\tenshf$ transform covariantly, as in\rfq{2.7} and\rfq{2.8b}. 

Next we consider a \df\ of the target space:
\numeq{2.9}{
\delta_\fkf \vX  = - \fkf (\vX) \ .
}
In contrast to the previous relabelings, this transformation is dynamical,
deforming the fluid configuration. Quantities
\numeq{2.8n}{
C_n (\vX) = \frac1{2^n n!}  
\eps_{i_1j_1\cdots
i_n j_n} \{X^{i_1}, X^{j_1}\}\cdots \{X^{i_n}, X^{j_n}\}
}
which are defined in $d=2n$~dimensions, respond to the transformation\rfq{2.9} in a
noteworthy fashion. One verifies that
\numeq{2.9n}{
\delta_\fkf C_n (\vX) =  -\grad\cdot\fkf(\vX)C_n(\vX)
}
so that transverse (volume preserving) target-space \df s leave $C_n$ invariant.
Eq.\rfq{2.9n} is most easily established by recognizing that
\numeq{2.10n}{
C_n (\vX) = \text{Pfaff} \{X^i, X^j\} 
= \det\nolimits^{1/2} \{X^i, X^j\}  
=\det\nolimits^{1/2} \vth \det \frac{\pr X^i}{\pr x^j}\ .
}
The significance of these transformations is evident from\rfq{1.17}, which shows that
$1/\rho(\vcr) = C_n(\vX)\bigr|_{\vx=\vchi(\vcr)}$ when $\det\nolimits^{1/2} \vth$ is
identified with
$1/\rho_0$. 
 The transformation law for~$\rho$ under
transverse target space
\df s becomes
\numeq{2.11}{
\delta_\fkf \rho(\vcr) =  \fkf \cdot \grad \rho(\vcr)\ .  
}
It follows that 
this transformation
leaves  invariant 
 all terms in the Lagrangian that depend only
on~$\rho$ [like $L_I$ in\rfq{1.23}]. 

When we restrict the transverse, target-space \df s to those that also leave $\vth$
invariant, i.e.,\rfq{3.3} (of course in two dimensions this is not a restriction), 
further quantities are left invariant. These are constructed as in\rfq{2.8n}, but with any
number of brackets $\{ X^i , X^j \}$ replaced by~$\tn ij$. 

It is interesting to combine the \df\ of the parameter space with that of the target space,
for a simultaneous transformation on both spaces. To this end we chose the form of the
target space transformation to coincide with that of the  reparameterization/relabeling
transformation.
\numeq{2.12}{
\fk^i(\vX) =  \tn ij\frac{\pr \phi(\vX)}{\pr X^j} \ . 
} 
As we shall show below,
this results in a gauge-covariant coordinate transformation on the vector
potential~$\vha$, once a further gauge transformation is carried out. Thus we consider 
$\Delta \equiv \delta_\phi + \delta_\fk$, 
\numeq{2.13}{
\Delta X^i = \{X^i, \phi(\vx)\} - \tn ij\frac{\pr \phi(\vX)}{\pr X^j} \ .  
}
[Note that any deviation of $\fkf^i (\vX)$ from $\tn ij\, {\pr\phi(\vX)}/{\pr X^j} $ may be
attributed to~$\phi$, and can be removed by a further gauge transformation.]
 However, covariance is not preserved in\rfq{2.13}:
$\vX$ on the left is covariant, but on the right in the Poisson bracket there occurs
$\phi(\vx)$, which is not covariant. The defect may be remedied by combining
$\Delta\vX$ with a further gauge transformation
\numeq{2.14}{
\delta_{\textrm{gauge}} \vX =   \{\vX, \phi(\vX) - \phi(\vx)\}
}
so that in $\Delta + \delta_{\textrm{gauge}} \equiv \hat\delta$ we have a covariant
transformation rule: 
\numeq{2.15}{
\hat\delta X^i =   \{X^i, \phi(\vX)\} - \tn ij \frac{\pr\phi(\vX)}{\pr X^j}
}
which in turn implies that $\vha$ transforms as 
\numeq{2.16}{
\hat\delta \ha_i =  \on ij \{X^j, \phi(\vX)\} - \frac{\pr\phi(\vX)}{\pr X^i}\ .
}

To recognize this transformation more clearly, we present it as
\begin{subequations}\label{2.17}
\numeq{2.17a}{
\hat\delta \ha_i =  \on ij \{X^j, X^k\} \frac{\pr\phi(\vX)}{\pr X^k} - \frac{\pr\phi(\vX)}{\pr
X^i}}
and use\rfq{2.4a} to find
\numeq{2.17b}{
\hat\delta \ha_i =  \tn k\ell \hat F_{i\ell} \frac{\pr\phi(\vX)}{\pr X^k} 
  = f^k (\vX)\hat F_{ki}\ .
 }
\end{subequations}

Note that in the final expression\rfq{2.17b} the response of $\vha$ is entirely covariant:
it involves the covariant curvature $\tenshf$ and the diffeomorphism function~$\vf$
evaluated on the covariant argument~$\vX$. This expression is precisely the
gauge-covariant \ct, which was previously derived\citer2, in a setting that differs from
the present in several ways. First,\citer2 dealt with a noncommuting theory, but we shall
presently extend the above to the noncommuting case. Second, in\citer2 the
transformation was not required to leave $\tensth$ invariant. With transformations that
change~$\tensth$, $\hat\delta$ includes the contribution $-(L_f\tn ij)\,\pr/\pr \tn ij$, and
the action of the $\tensth$ derivative is evaluated by the \sw\ equation\citer3. Finally, in
order to simplify ordering problems, $\vf$ was restricted to be at most linear in its
argument, and the noncommutative formula corresponding to\rfq{2.17b} involved a
(star)~anticommutator.

\subsection{Noncommuting theory with star products (even dimensions)}\label{s2ssC}

The above development may be taken over directly into  a noncommutative field theory
by replacing Poisson brackets by $-i$~times~(star) commutators, so that\rfq{2.2b} goes
over into\rfq{1.1}. Eq.\rfq{1.14} remains and\rfq{2.4a},\rfq{2.5} become
\begin{align}
[X^i,X^j]_\star &= i\tn ij + i\tn ik \tn j\ell \hat F_{k\ell}\label{2.18}\\
\hat F_{ij} &= \pa{x^i} \ha_j - \pa{x^j} \ha_i - i[\ha_i, \ha_j]_\star\ .\label{2.19}
\end{align}

The covariant transformation rules\rfq{2.15} and\rfq{2.16} may be used in the
noncommutative context, provided a sensible ordering prescription is set for~$\phi(\vX)$.
This we do as follows. Define
\begin{subequations}\label{2.20}
\numeq{2.20a}{
\Phi = \int \rd x \phi(\vX)
}
where $\phi(\vX)$ is a series of (star) powers of~$\vX$:
\numeq{2.20b}{
\phi(\vX) = c + c_i X^i + \fract12 c_{ij}X^i\star X^j + \fract13 c_{ijk} X^i\star X^j \star X^k +
\cdots\cdot
 }
\end{subequations}
[We are not concerned about convergence of the integral\rfq{2.20a}, 
since we are interested in local
quantities like \rfq{2.20b} or \rfq{2.23} below.]
The integration over $\vx$ (the argument of~$\vX$) ensures that $\Phi$ is invariant (in an
operator formalism the integral becomes the trace of the operators). The $c$-coefficients
in\rfq{2.20b} enjoy cyclic invariance (so that $\Phi$ and $\phi$ possess the same number
of free parameters).  Also we require $\phi$ to be Hermitian. [This ensures, e.g., that
$c_{ij}$ is real symmetric; that $\Re  c_{ijk}$ is entirely symmetric and that $\Im  c_{ijk}$ is
entirely antisymmetric (which is impossible in two dimensions).] Then\rfq{2.15}
and\rfq{2.16} become
 \begin{align}
\hat\delta X^i &= -i[X^i, \phi(\vX)]_\star - \tn ij \frac{\delta\Phi}{\delta
X^j}\label{2.21}\\[0.5ex]
\hat\delta \ha_i &= -i\on ij[X^j, \phi(\vX)]_\star  -  \frac{\delta\Phi}{\delta X^i}\label{2.22}
\end{align}
where now the last entries employ a functional derivative:
\numeq{2.23}{
\frac{\delta\Phi}{\delta X^i} = c_i +  c_{ij} X^j + c_{ijk}  X^j \star X^k +
\cdots\cdot
 }

In two dimensions, 
the ordering prescription\rfq{2.20} and its consequence\rfq{2.23}
preserve the invariance of the $[X^i, X^j]_\star$ commutator against the target space \df\
[last term in\rfq{2.21}]. Thereby a property of the classical Poisson bracket
[c.f.\rfq{2.9n} at $n=1$] is maintained in the noncommuting theory.

With $\phi(\vX)$ at most quadratic in~$\vX$ ($\vf$ at most linear) as in\citer2, one
readily verifies the result in that paper
\numeq{2.24}{
\delta \ha_i = \fract12 \bigl\{ f^j (\vX)\star \hat F_{ji}  +   \hat F_{ji} \star f^j
(\vX)
\bigr\}\ .
 }
But with more general $\phi$ ($\vf$ containing quadratic and higher powers) there
arise further reordering terms. 

\subsection{Commuting and noncommuting transformations (odd
dimensions)}\label{s2ssE}

In odd dimensions, with the $\vth$-preserving transformation function given
by\rfq{1.13}, the relabeling transformation on 
the base space is
\numeq{3.13}{\begin{aligned}
\delta_\phi \vX\tvx &=  \tn ij \pa{x^j} \phi\tvx \pa{x^i} \vX\tvx  
    + f(t) \pa t \vX\tvx \\[1ex]
  &= \bigl\{\vX\tvx, \phi\tvx  \bigr\} + f(t)\dot{\vX}\tvx   \ .
\end{aligned}
}
The fluid coordinate $\vX$ has components only in the spatial directions. Here the Poisson
bracket is defined with the nonsingular~$\tn ij$. 

For the target space \df\ we again take the formula\rfq{2.12}, so that the combined,
noncovariant transformation $\Delta\equiv\delta_\phi+\delta_\fk$ reads 
\numeq{3.14}{
\Delta  X^i = \bigl\{ X^i , \phi\tvx\bigr\}
 +   f(t)\dot X^i\tvx - \tn ij \frac{\pr\phi\tvX}{\pr X^j}\ . 
}
This is modified by the gauge transformation
\numeq{3.15}{
\delta_{\textrm{gauge}} \vX = \bigl\{ \vX, \phi\tvX - \phi\tvx \bigr\}
 -   \bigl\{ \vX, f(t)\han 0 \tvx \bigr\} 
}
resulting in the covariant transformation $\Delta + \delta_{\textrm{gauge}} 
\equiv \hat\delta$:
\numeq{3.16}{
\hat\delta  X^i = \bigl\{ X^i , \phi\tvX\bigr\}
  - \tn ij \frac{\pr\phi\tvX}{\pr X^j} + f(t) D{X^i} \ . 
}
Here $D{X^i} = \dot X^i + \{ \ha_0, X^i \}$, where $\ha_0$ is a connection introduced to
render the time derivative covariant against time-dependent gauge transformations,
generated by $\phi$. This is achieved when the gauge transformation law for $\ha_0$ is 
\numeq{3.17}{
\delta_\phi \ha_0=  \dot\phi +   \bigl\{ \han0,  \phi \bigr\} \ .
}
The spatial components of the vector potential are introduced as before in\rfq{1.14}
\numeq{3.20}{
  D{X^i} = \tn ij \Bigl( \skew5\dot{\hat A}_j  -\pr_j \han 0 + \{\han 0, \han j\} \Bigr)
 = \tn ij \hat F_{0j}\ .
}
The covariant transformation law of $\vha$ follows from\rfq{2.19},\rfq{3.16},
and\rfq{3.20}: 
\numeq{3.21}{\begin{aligned}
\hat\delta \han i &= \on ij \bigl\{ X^j , \phi\tvX\bigr\}
  -  \frac{\pr\phi\tvX}{\pr X^i} +  \on ij f(t) D{X^j} \\[1ex]
  &= f^j \tvX \hat F_{ji} + f(t) \hat F_{0i} = f^\mu \tvX \hat F_{\mu i}\ .
\end{aligned}
}
It remains to fix the transformation law of $\han 0$. This requires specifying
$\delta_\fk \han 0$. Since 
\numeq{2.35}{
\delta_\fk \han i =  -  \frac{\pr\phi\tvX}{\pr X^i}
} it is natural
to take 
\numeq{2.36}{
\delta_\fk \han 0 =  -  \frac{\pr\phi\tvX}{\pr t}
}
(The derivative acts on the first argument only.)
Thus we have from\rfq{3.17} and\rfq{2.36}
\numeq{2.37}{
\Delta \han 0 =  \frac{\pr\phi\tvx}{\pr t} + \{\han 0, \phi\tvx\} -  
\frac{\pr\phi\tvX}{\pr t}\ . 
}
After adding to this a gauge transformation generated by
$\phi\tvX - \phi\tvx$ we are left with
\numeq{3.22}{
\begin{aligned}
  \hat\delta \han0  &= \frac{\pr\phi\tvX}{\pr X^i} \frac{\pr X^i}{\pr t} + \{\han 0,
 \phi\tvx\}\\
 &= \frac{\pr\phi}{\pr X^i} \rD{X^i} =  f^i \tvX \hat F_{i0} =  f^\mu \tvX \hat F_{\mu 0}\ .
\end{aligned}
}
Eqs.\rfq{3.21} and\rfq{3.22} coincide with the formula obtained in a conventional
commuting gauge theory\citer2.

Similar results follow within the noncommuting formalism, once the now familiar ordering
prescription is given for $\phi\tvX$ and $\Phi = \int \rd x \phi\tvX$. In the \nc\
formalism\rfq{3.21} and\rfq{3.22} are regained, up to reordering terms.

\subsection{\sw\ map}\label{s2ssD}

To construct the \sw\ map in two Euclidean dimensions, we (temporarily) introduce a
time dependence in the fluid variables (but not into the \df\ functions -- only spatial
variables are transformed) and 
observe from\rfq{1.20} that $(\rho,\rho\vv)$ form a conserved 3-vector~$j^\alpha$ 
[also true  in the noncommuting theory when an ordered definition for
$\delta(\vX\tvx - \vcr)$ is given -- this will be provided below].
Therefore, the dual of~$j^\alpha$, $\eps_{\mn\alpha} j^\alpha$,  
satisfies a Bianchi identity and can be presented as the curl of a potential, apart from
additive and multiplicative constants:
\begin{gather}
 \eps_{\mn\alpha} j^\alpha \propto F_\mn + \text{constant}\label{2.25}\\
F_\mn = \pr_\mu A_\nu - \pr_\nu A_\mu\ .\label{2.26}
\end{gather}
Note $j^\alpha$, $F_\mn$, $A_\mu$ are ordinary functions, even in the noncommuting
setting, since the noncommuting variables~$\vX$ are integrands (in an operator
formalism, their trace is involved). In particular, the spatial tensor is determined
by~$\rho$:
\numeq{2.27}{
\pa{r^i} A_j (\vcr)  - \pa{r^j} A_i (\vcr) = F_{ij}(\vcr) 
   = - \eps_{ij} (\rho -\rho_0)
 =  - \eps_{ij} \rho_0 \Bigl(\int \rd x \delta\bigl( \vX(\vx) - \vcr\bigr) - 1\Bigr)\ . 
 }
(The time dependence is now suppressed.)
$\vX$ contains $\vha$, as in\rfq{1.14}, but it is (noncommuting) gauge covariant, so the
integral in\rfq{2.27} is (noncommuting) gauge invariant. Therefore,\rfq{2.27} serves to
define an (inverse) \sw\ map between the noncommuting (hatted) and commuting
(unhatted) variables.
The additive ($\eps_{ij} \rho_0 $) and multiplicative ($-1$) constants are fixed by
requiring agreement at small $\vha$.
 It still remains to give a proper ordering to the $\delta$-function
containing~$\vX$. This we do by a Fourier transform prescription:
\numeq{2.28}{
\int \rd r e^{i\vk\cdot\vcr} F_{ij}(\vcr)
   = -\eps_{ij}\rho_0 \int \rd x \bigl( e^{i\vk\cdot\vX(\vx)}_\star - e^{i\vk\cdot\vx}\bigr)
 }
and the ordering (Weyl ordering) is defined by the expansion of the exponential in (star
product) powers. 

When the exponential $e^{i\vk\cdot\vX}_\star\equiv 
1+ i \vk \cdot \vX - \frac12  \vk \cdot \vX \star \vk \cdot \vX + \cdots$ 
is written explicitly in terms of~$\vha$:
$\exp_\star i( k_i x^i + \theta k_i \en ij \ha_j)$, factoring the exponential 
into $e^{i\vk\cdot\vx}$~times another factor involves the Baker-Hausdorff 
lemma, and leads
to an open Wilson line integral\citer6. In that form\rfq{2.28} is seen to 
coincide with the known solution to the
\sw\ map\citer4, which is now also recognized as nothing but an instance of the
Lagrange$\,\to\,$Euler map of fluid mechanics. (See also \citer7.)
 
To construct the \sw\ map in (2+1)-dimensional space-time we consider
the conserved current, defined in\rfq{1.16} and\rfq{1.19}, except that now the time
dependence is retained  throughout and the 
derivative is gauged with $\han0$:
\numeq{3.23}{
  \vj\tvr = \int\rd x \bigl(\dot \vX + \{\han0, \vX\}\bigr) \delta (\vX-\vcr)\ .
}
The operator ordering is prescribed in momentum space with the exponential (Weyl)
ordering and\rfq{3.23} in the noncommuting theory becomes
\numeq{3.24}{
\vj \tvk  \equiv
\int \rd r e^{i\vk\cdot\vcr} \vj \tvr = \int\rd x e^{i\vk\cdot\vX}_\star \bigl(\dot{\vX} - 
i[\han 0,
\vX]_\star\bigr)
\ . }
Note that the commutator does not contribute to current conservation because it is
separately transverse:
\numeq{3.25}{
 \int\rd x e^{i\vk\cdot\vX}_\star \bigl[\han 0, \vk\cdot\vX\bigr]_\star =
\int\rd x \han 0  \bigl[\vk\cdot\vX, e^{i\vk\cdot\vX}_\star\bigr]_\star = 0 \ .
}
Therefore the 3-current is conserved as before. Its dual, $\eps_{\mn\alpha} j^\alpha$
satisfies the Bianchi identity, so the \sw\ mapping reads
\numeq{3.26}{\begin{aligned}
\int\rd r e^{i\vk\cdot\vcr}  \bigl(1 - \fract12 \tn ij F_{ij}\bigr) &= 
     \int\rd x e^{i\vk\cdot\vX}_\star \\[1ex]
\int\rd r e^{i\vk\cdot\vcr} F_{0i} &= 
   \on ij \int\rd x e^{i\vk\cdot\vX}_\star \bigl(\dot X^j - i[\han0, X^j]_*\bigr)\\
   &= \int\rd x e^{i\vk\cdot\vX}_\star \hat F_{0i} \ .
\end{aligned}
}

Formulas\rfq{2.28} and\rfq{3.26} may be verified by comparison with the explicit
$\mathcal O(\theta)$
\sw\ map, which  for field strengths reads
\numeq{3.27}{
F_\mn = \hat F_\mn - \tn \alpha\beta (\hat F_{\alpha\mu} \hat F_{\beta\nu} -
        \han \alpha \pr_\beta \hat F_\mn)\ .
}
Upon setting $\tn \alpha0 = 0$, $\tn ij = \theta \en ij$ and 
\numeq{3.28}{
e^{i k_i (x^i +\tn ij\han j)}_\star = e^{i\vk\cdot\vx} \Bigl(1 + i\theta k_i \en ij \han j 
- \fract12 \theta^2 k_i k_m \en ij \en mn \han j \han n \Bigr)
}
it is recognized that\rfq{2.28} and\rfq{3.26} reproduce\rfq{3.27}.

\section{\sw\  Map in Higher Dimensions}\label{s3}

\subsection{Even dimensions}\label{s3ssA}

In dimensions higher than three the correspondence between the Bianchi
identity and the conservation of particle current is lost. The derivation
of the \sw\ map calls for higher conserved currents, whose duals are
two-forms. 

The introduction of such currents can be motivated by starting again from
the commutative particle density $\rho$ as expressed in \rfq{1.16b} and
its inverse $\rho^{-1}$ as expressed in \rfq{1.17}. Their product
\numeq{aaa}{
1 =  \int \rd x  \delta \bigl(\vX - \vcr\bigr)
\det \frac{\pr X^i (\vx)}{\pr x^j} 
}
is independent of the fluid profile $\vX (\vx)$ and constitues a topological
invariant. The Jacobian determinant in the above can be expressed in
terms of the square-root determinant (Pfaffian) of the antisymmetric 
matrix $\{ X^j , X^k \}$:
\numeq{bbb}{
1 = \frac{\rho_0}{2^n n!} \int \rd x  \delta \bigl(\vX - \vcr\bigr)
\epsilon_{i_1 , j_1 , \dots ,i_n , j_n} \{ X^{i_1} , X^{j_1} \}
\cdots \{ X^{i_n} , X^{j_n} \} = \rho_0 \int \rd x 
\delta \bigl(\vX - \vcr\bigr)  C_n (\vX )
}
where, in analogy with the 2-dimensional case, we identified 
${\rm Pfaff} ({\vth})$ with $1/\rho_0$. Removing all
$n$ Poisson brackets from the above recovers the full density $\rho$.
The removal of a single Poisson bracket $\{ X^i , X^j \}$, then,
produces a sort of residual density $\rho_{ij}$ in the corresponding
dimensions, which becomes a candidate for the \sw\ commutative field
strength:
\numeq{bbc}{
\rho_{ij} = \frac{\rho_0}{2^{n-1} (n-1)!} \int \rd x  \delta 
\bigl(\vX - \vcr\bigr) \, \epsilon_{i,j,i_2,j_2, \dots ,i_n ,j_n} 
\{ X^{i_2} , X^{j_2} \} \cdots \{ X^{i_n} , X^{j_n} \}
}
The current dual to $\rho_{ij}$, in momentum space,
\numeq{ccc}{
J^{j_1 \dots j_{2n-2}} = \frac{\rho_0}{2^{n-1} (n-1)!}
\int \rd x  e^{i \vk \cdot \vX} \{ X^{[ j_1} , X^{j_2} \}
\cdots \{ X^{j_{2n-3}} , X^{j_{2n-2} ]} \}
}
(the indices are fully antisymmetrized) is gauge-invariant and conserved, 
ensuring that $\rho_{ij}$ satisfies the Bianchi identity. 

The corresponding current in the noncommutative case can be written by 
turning products
into star-products and Poisson brackets into ($-i$ times) star-commutators. 
The ordering
of the exponential and other factors above has to be fixed in a way
which ensures that the obtained current is conserved. Various such orderings
are possible. For definiteness, we pick the ordering corresponding to the
choice made in\citer4 :
\begin{eqnarray}
J^{j_1 \dots j_{2n-2}} = \frac{\rho_0}{(2i)^{n-1}}
\int \rd x && \int_0^1 ds_1 \cdots \int_0^1 ds_{n-1}
\delta \left( 1- \sum_{i=1}^{n-1} s_i \right) \nonumber \\
&&e_*^{i s_1 \vk \cdot \vX } \bigl[ X^{[ j_1} , X^{j_2} \bigr]_*  * 
\cdots e_*^{i s_{n-1} \vk \cdot \vX } * \bigl[ X^{j_{2n-3}} , 
X^{j_{2n-2}] } \bigr]_*
\label{ddd}
\end{eqnarray}
This corresponds to Weyl-ordering the exponential and distributing it
in all possible ways between the different commutators. Note that the
volume of the $s_i$-integration space reproduces the factor $1/(n-1)!$
present in \rfq{ccc}.

To express compactly the above and facilitate the upcoming derivations, 
we introduce antisymmetric tensor notation. We define the
basis one-tensors ${\rm v}_j$ representing the derivative vector field
$\partial_j$, and corresponding one-forms $\rd x^j$. We consider the 
fundamental one-tensor $\rX$ and the one-form $\rk$ 
\numeq{eee}{
\rX = X^j {\rm v}_j ~,~~~~ \rk = k_j {\rd x}^j
}
All tensor products will be understood as antisymmetric, i.e.,
\numeq{fff}{
{\rm v}_j {\rm v}_k \equiv \frac{1}{2} \left(
{\rm v}_j \wedge {\rm v}_k - {\rm v}_k \wedge {\rm v}_j \right)
}
etc.,
which amounts to considering ${\rm v}_j$ and $\rd x^k$ as anticommuting 
quantities. Scalar products are given by the standard contraction 
\numeq{ffg}{
{\rm v}_j \cdot {\rd x}^k = \delta_k^j
}
We also revert to operator notation, dispensing with star products
and writing $\Tr$ for $\rho_0 \int \rd x$. 
Finally, we simply write $\int_{(n-1)}$
for the ($n-1$)-dimensional $s_i$-integration
\numeq{fgg}{
\int_{(n-1)} \equiv \frac{1}{(2i)^{n-1}} \int_0^1 ds_1 \cdots
\int_0^1 ds_{n-1} \delta \left( 1- \sum_{i=1}^{n-1} s_i \right)
}
Overall, the current in \rfq{ddd} is written as the rank-($2n-2$)
antisymmetric tensor $\rJ$
\numeq{ggg}{
\rJ = \Tr  \int_{(n-1)}
e^{i s_1 \rk \cdot \rX } \rX \rX \cdots e^{i s_{n-1} \rk \cdot \rX } 
\rX \rX
}
and its conservation is expressed by the contraction $\rk \cdot \rJ =0$.
The contraction of $\rk$ acts on each $\rX$ in a graded fashion.
Using cyclicity of trace and invariance under relabeling the $s_i$, 
this becomes
\numeq{hhh}{
\rk \cdot \rJ = (n-1) \Tr  \int_{(n-1)}
e^{i s_1 \rk \cdot \rX } \bigl[ \rk \cdot \rX , \rX \bigr]
e^{i s_2 \rk \cdot \rX } \rX \rX \cdots e^{i s_{n-1} \rk \cdot \rX } \rX \rX
}
Using the identity
\numeq{iii}{
\bigl[ e^{i s \rk \cdot \rX} , \rX \bigr] =
\int_0^s  ds_1 e^{i s_1 \rk \cdot \rX } \bigl[ i \rk \cdot \rX , \rX \bigr]
e^{i (s - s_1 ) \rk \cdot \rX }
}
we can absorb the $s_1$-integration in \rfq{hhh} and bring it in the form
\numeq{jjj}{
\rk \cdot \rJ = -\frac{1}{2} \Tr  \int_{(n-2)}
\bigl[ e^{i s_2 \rk \cdot \rX} , \rX \bigr] 
\rX \rX \cdots e^{i s_{n-1} \rk \cdot \rX } \rX \rX
}
Finally, using once more the cyclicity of trace, we see that the above
contraction vanishes. This proves that the tensor $\rJ$ is conserved and,
as a consequence, its dual $\rho_{jk}$ satisfies the Bianchi identity.
As in the 2-dimensional case, we put
\numeq{kkk}{
F_{jk} (\vk ) = \rho_{jk} (\vk) - \omega_{jk} \delta (\vk)
}
and recover the commuting Abelian field strength, which can, in turn, be
expressed in terms of a (commutative) abelian potential $A_j$.

In the above manipulations we freely used cyclicity of trace. In general
this is dangerous, since the commuted operators may not be trace class. 
Assuming, however, that $\vX$ becomes asymptotically $\vx$ for large distances,
the presence of the exponentials in the integrand
ensures that this operation is permissible.

As mentioned previously, the fully symmetric ordering is not the only
one that leads to an admissible $\rho_{jk}$. As an example, in the
lowest-dimensional nontrivial case $d=4$ we can alter the ordering
by splitting the commutator as
\numeq{lll}{
J^{jk} = \frac{1}{2i} \Tr \left\{ e^{i \vk \cdot \vX } \bigl[ X^j , X^k \bigr]
\right\} ~ \to ~
J_f^{jk} = -i \Tr \int_0^1 ds f(s) e^{i s \vk \cdot \vX } X^j 
e^{i (1-s) \vk \cdot \vX } X^k 
}
If $f(s) = - f(1-s)$ the above will be antisymmetric in ($j,k$) and 
conserved, as can explicitly be verified. Further, if $f(s)$ satisfies
\numeq{mmm}{
\int_0^1 ds (2s-1) f(s) = 1
}
then \rfq{lll} will also have the correct commutative limit. We obtain an
infinity of solutions depending on a function of one variable $f(s)$.
This arbitrariness reflects the fact that the \sw\ equations are not
integrable and therefore the solution for $\vth = 0$ depends on the
path in the $\vth$-space taken for integrating the equations. For
$d=4$ the parameter space is a plane and the
path from a given $\vth$ to $\vth =0$ on the plane
can be parametrized by a function of a single variable, just like
$J_f^{jk}$. The various solutions are related through field redefinitions.

\subsection{Odd  dimensions}\label{s3ssB}

The situation in odd dimensions differs in that we need to specify
separately the components of the conserved current in the commutative
and noncommutative directions. For $d=2n+1$ the current is of rank
$2n-1$ and it can be constructed by a procedure analogous to the
even-dimensional case: We start from the expression for the total
particle current $j^\mu$ \rfq{1.16} and \rfq{1.19} and introduce 
$2n-2$ commutators, one less than the number which would fully saturate it 
to $(1, \vv )$. The temporal components $J^{0 j_1 \dots j_{2n-2}}$ can be
expressed as a rank-$(2n-2)$ antisymmetric spatial tensor $\rJ^0$,
while the spatial components $J^{j_0 j_1 \dots j_{2n-2}}$ can be expressed
as a rank-$(2n-1)$ antisymmetric tensor $\rJ$. Their fully
ordered expressions are
\numeq{nnn}{
\rJ^0 = \frac{1}{n-1} \Tr  \int_{(n-1)}
e^{i s_1 \rk \cdot \rX } \rX \rX \cdots e^{i s_{n-1} \rk \cdot \rX } \rX \rX
}
\numeq{ooo}{
\rJ = \Tr  \int_{(n)} e^{i s_0 \rk \cdot \rX} D \rX
e^{i s_1 \rk \cdot \rX } \rX \rX \cdots e^{i s_{n-1} \rk \cdot \rX } \rX \rX
}
The above expressions can be unified by introducing a temporal component
for the field $X^\mu$, namely $X^0 \equiv t$ (which is obviously commutative),
and extending the one-tensor $\rX$ also to include $X^0 {\rm v}_0$. Further,
we can Fourier transform in time and define $\rk = k_\mu {\rd x}^\mu$ to
also include the frequency $k_0$. Then the corresponding (space-time)
($2n-1$)-tensor $\rJ$ acquires the form
\numeq{oop}{
\rJ = \int dt ~\Tr  \int_{(n)} e^{i s_1 \rk \cdot \rX} \, D \rX \,
e^{i s_2 \rk \cdot \rX } \rX \rX \dots e^{i s_n \rk \cdot \rX } \rX \rX
}
$X^0$ is absent in $\rX \rX$ and, since $D X^0 = 1$, only $s_0 + s_1$ appears 
in the temporal component of $\rJ$; integrating over $s_1$ reproduces
the factor $1/(n-1)$ appearing in \rfq{nnn}.

The above current is obviously gauge invariant. We shall prove that it
is also conserved, that is, it satisfies $\rk \cdot \rJ = 0$.
The contraction is
\vfill
\eject
\begin{eqnarray}
\rk \cdot \rJ = \int dt ~\Tr  \int_{(n)} && \Bigl\{
e^{i s_1 \rk \cdot \rX} \, \rk \cdot D\rX \,
e^{i s_2 \rk \cdot \rX } \rX \rX \cdots e^{i s_n \rk \cdot \rX } \rX \rX
\nonumber \\
&-& \sum_{m=2}^n 
e^{i s_1 \rk \cdot \rX} \, D\rX \, \rX \rX \cdots
e^{i s_m \rk \cdot \rX } \bigl[ \rk \cdot \rX , \rX \bigr] 
e^{i s_{m+1} \rk \cdot \rX } \cdots \rX \rX \Bigr\}
\end{eqnarray}
(with $s_{n+1} =0$). By formula \rfq{iii} and a similar one for the
covariant time derivative, the above can be rewritten as
\begin{eqnarray}
\rk \cdot \rJ = \int dt ~\Tr  \int_{(n-1)} && \Bigl\{
De^{i s_1 \rk \cdot \rX} \,
e^{i s_2 \rk \cdot \rX } \rX \rX \cdots e^{i s_{n-1} \rk \cdot \rX } \rX \rX
 \nonumber \\
&-& \sum_{m=2}^{n-1} 
e^{i s_1 \rk \cdot \rX} \, D\rX \, \rX \rX \cdots
\bigl[ e^{i s_m \rk \cdot \rX } , \rX \bigr] 
\rX \rX \cdots \rX \rX \Bigr\}
\end{eqnarray}
Due to the cyclicity of trace, the sum above telescopes and only the
first term of the $m=2$ commutator and the second term of the $m=n-1$
commutator survive. Altogether we obtain
\begin{eqnarray}
\rk \cdot \rJ &=& \int dt ~\Tr  \int_{(n-1)} \left(
D e^{i s_1 \rk \cdot \rX} + D \rX \rX + \rX D \rX \right)
e^{i s_2 \rk \cdot \rX } \rX \rX \cdots e^{i s_{n-1} \rk \cdot \rX } \rX \rX
\nonumber \\
&=& \int dt ~\Tr  \int_{(n-1)} D \left( e^{i s_1 \rk \cdot \rX} \rX \rX \right) 
\cdots e^{i s_{n-1} \rk \cdot \rX } \rX \rX \nonumber \\
&=& \int dt ~\Tr  \int_{(n-1)} \frac{1}{n-1} D \left(
e^{i s_1 \rk \cdot \rX} \rX \rX
\cdots e^{i s_{n-1} \rk \cdot \rX } \rX \rX \right) \\
&=& \int dt ~\frac{d}{dt} \Tr  \int_{(n-1)} \frac{1}{n-1}
e^{i s_1 \rk \cdot \rX} \rX \rX
\cdots e^{i s_{n-1} \rk \cdot \rX } \rX \rX \nonumber \\
&=& 0 \nonumber
\end{eqnarray}
which proves the conservation of $\rJ$. Its dual $\rho_\mn$ satisfies
the ($2n+1$)-dimensional Bianchi identity and can be used to define the
commutative Abelian field stength
\begin{eqnarray}
F_{ij} (\vk ) &=& \rho_{ij} (\vk) - \omega_{ij} \delta (\vk) \\
F_{0i} (\vk ) &=& \rho_{0i} (\vk )
\end{eqnarray}

In the above we gave separate derivations of the \sw\ map for even
and odd dimensions. The two can be unified by demonstrating that each
case can be obtained as a dimensional reduction of the other in one
more dimension. This is treated in the next section.
 
\subsection{Dimensional reduction}\label{s3ssC}

It is quite straightforward to see that the even dimensional \sw\ map
is obtained from the $d=2n+1$ map by dimensional reduction.
We assume a time-independent configuration in which $X^j$ 
($j=1,\dots ,2n$) do not depend on $t$ and $A_0$ vanishes.
In this case $D \rX$ vanishes and so does
$\rJ$ in\rfq{ooo}; only the component $\rJ^0$ 
in\rfq{nnn} survives, reproducing the $2n$-dimensional solution.

The reduction from a fully noncommutative $d=2n+2$ case to the $d=2n+1$
case is only slightly subtler. For concreteness, we shall take 
$t \equiv x^0$ to be canonically conjugate to the last dimension, 
call it $z \equiv x^{2n+1}$, which will be reduced; that is, 
\numeq{separ}{
[ t , z ] = i \theta_0 ~~(\theta_0 = \theta^{0,2n+1})~,~~~
[ t , x^i ] = [ z , x^i ] = 0 ~~(i=1,\dots ,2n)
}
This can always be achieved with an orthogonal rotation of the $x^\mu$.
The reduced configuration consists of taking all fluid coordinates other 
than $X^{2n+1}$ to be independent of $x^{2n+1}$ and, further, the gauge
potential corresponding to $z=x^{2n+1}$ to vanish. Specifically,
\begin{eqnarray}
X^i &=& X^i ( \vx , t) \\
X^0 &=& t \\
X^{2n+1} &=& z + \theta_0 A_0 ( \vx , t) 
\end{eqnarray}
With this choice the corresponding field strengths become
\begin{align}
[X^i,X^j] &= i\tn ij + i\tn ik \tn j\ell \hat F_{k\ell}\label{fij}\\
[ X^i , X^0 ] &= 0 \\
[ X^i , X^{2n+1} ] &= i \theta_0 ( D_0 X^i - i [X^i , A_0 ] ) 
= i\theta^{ij} \theta^{2n+1,0} \hat F_{j0}
\end{align}
with $\hat F_{\mu \nu}$ ($\mu,\nu=0,\dots,2n$) the field strength of
a noncommutative $d=2n+1$ theory.

The corresponding $d=2n+2$ \sw\ map reduces to the $d=2n+1$ map.
Indeed, the current $\rJ$ in\rfq{ggg}, now, is a rank-$2n$ antisymetric 
tensor. When all its indices are spatial ($1,\dots,2n$) it becomes a fully
saturated topological invariant, that is, a constant; this reproduces
a constant $\rho_{0,2n+1}$. When one of its indices is $0$ and
the rest are spatial it vanishes, leading to $\rho_{i,2n+1} = 0$.
When one of its indices is $2n+1$ and the rest are spatial it reproduces
expression \rfq{ooo}. Finally, when two of its indices are $0,2n+1$
and the rest are spatial it reproduces\rfq{nnn}, recovering the full
commuting ($2n+1$)-dimensional Abelian field strength. 

We stress that the above reductions are not the most general ones.
Indeed, mere invariance of the fluid configuration with respect to
translations in the extra dimension does not require the vanishing of
the gauge field in the corresponding direction. This means that we
could choose $X^0 = t + H (\vx ,t)$ (instead of $X^0 = t$) in both
$d=2n+1$ and $d=2n+2$. The corresponding reduced theory contains an
extra Higgs scalar in the adjoint representation of the (noncomutative)
U(1) gauge group. Our \sw\ map in this situation reproduces, with
no extra effort, the space-time derivatives of a corresponding
commuting `Higgs' scalar.

We conclude by remarking that the above complete reduction scheme 
($2n+2 \to 2n+1 \to 2n \to \dots$) is reminiscent of the topological 
descent equations relevant to gauge anomalies. This may prove fruitful
in the analysis of noncommutative topological actions and the mapping
of topologically nontrivial configurations \citer8.

\subsection*{Acknowledgments}

The research is
supported in part by funds provided by the U.S. Department of Energy (D.O.E.)
under cooperative research agreements Nos.\ DE-FC02-94-ER40818 and 
DE-FG02-91-ER40676 and DOE-91ER40651-TASKB.


\begin{thebibliography}{99}
\frenchspacing

\bibitem{1} 
L.~Susskind,
arXiv:hep-th/0101029.


\bibitem{2} 
R.~Jackiw and S.~Y.~Pi,
Phys.\ Rev.\ Lett.\  {\bf 88}, 111603 (2002)
[arXiv:hep-th/0111122].


\bibitem{3} 
N.~Seiberg and E.~Witten,
JHEP {\bf 9909}, 032 (1999)
[arXiv:hep-th/9908142].


\bibitem{4} 
Y.~Okawa and H.~Ooguri,
Phys.\ Rev.\ D {\bf 64}, 046009 (2001)
[arXiv:hep-th/0104036].


\bibitem{5} 
For more extensive discussions see, e.g., V.~Arnold and B.~Khesin,
\emph{Topological Methods in Hydrodynamics} (Springer, Berlin, 1998);
R.~Jackiw, \emph{Lectures on Fluid Dynamics} (Springer, Berlin, 2002);
I.~Antoniou and G.~Pronko, hep-th/0106119.
 

\bibitem{6} 
J.~Madore, S.~Schraml, P.~Schupp and J.~Wess,
Eur.\ Phys.\ J.\ C {\bf 16}, 161 (2000)
[arXiv:hep-th/0001203];
B.~Jurco, S.~Schraml, P.~Schupp and J.~Wess,
Eur.\ Phys.\ J.\ C {\bf 17}, 521 (2000)
[arXiv:hep-th/0006246];
D.~J.~Gross, A.~Hashimoto and N.~Itzhaki,
Adv.\ Theor.\ Math.\ Phys.\  {\bf 4}, 893 (2000)
[arXiv:hep-th/0008075];
D. Bak, K. Lee, and J.-H.~Park,
Phys.\ Lett.\ B {\bf 501}, 305 (2001)
[arXiv:hep-th/0011244].


\bibitem{7}
P.~C.~Stichel,
Phys.\ Lett.\ B {\bf 526}, 399 (2002)
[arXiv:hep-th/0112025].

\bibitem{8}
A.~P.~Polychronakos, 
Annals Phys.\ (in press)
[arXiv:hep-th/0206013].

	 
\end{thebibliography}
\end{document}